\documentclass{moriond}

\def\be{\begin{equation}}
\def\ee{\end{equation}}
\def\bea{\begin{eqnarray}}
\def\eea{\end{eqnarray}}

\newcommand{\Photo}{}

\begin{document}
\vspace*{4cm}
\title{Top quark mass and cross section at ATLAS and CMS}

\author{Nils Faltermann on behalf of the ATLAS and CMS Collaborations}

\address{Karlsruhe Institute of Technology (KIT), Institute of Experimental Particle Physics,\\
  Wolfgang-Gaede-Str. 1, 76131 Karlsruhe, Germany}

\maketitle\abstracts{The top quark is the heaviest elementary particle known to date and therefore an important topic to study in the context of the standard model at the LHC. In this contribution the latest measurements of top quark production cross sections and the top quark mass at the LHC by the ATLAS and CMS Collaborations are presented.
}

\let\thefootnote\relax\footnote{Copyright 2024 CERN for the benefit of the ATLAS and CMS Collaborations. Reproduction of this article or parts of it is allowed as specified in the CC-BY-4.0 license}

\section{Introduction}
At the Large Hadron Collider (LHC), top quarks are predominantly produced as top quark-antiquark pairs ($\mathrm{t\overline{t}}$) via the strong interaction, but can also be produced as single top quarks via the electroweak interaction. Since the top quark almost always decays into a W boson and a bottom quark before any hadronization the final state is characterized by the subsequent decay of the W boson, which can either happen into a quark-antiquark pair or into a charged lepton and the corresponding neutrino. Due to their abundant production at the LHC and their clear experimental signature top quarks provide a unique tool for probing standard model (SM) parameters and search for deviations of theoretical predictions, which would be an indication for physics beyond the SM.

\section{Inclusive top quark cross sections}
\begin{figure}
\begin{minipage}{0.4\linewidth}
\centerline{\includegraphics[width=1\linewidth]{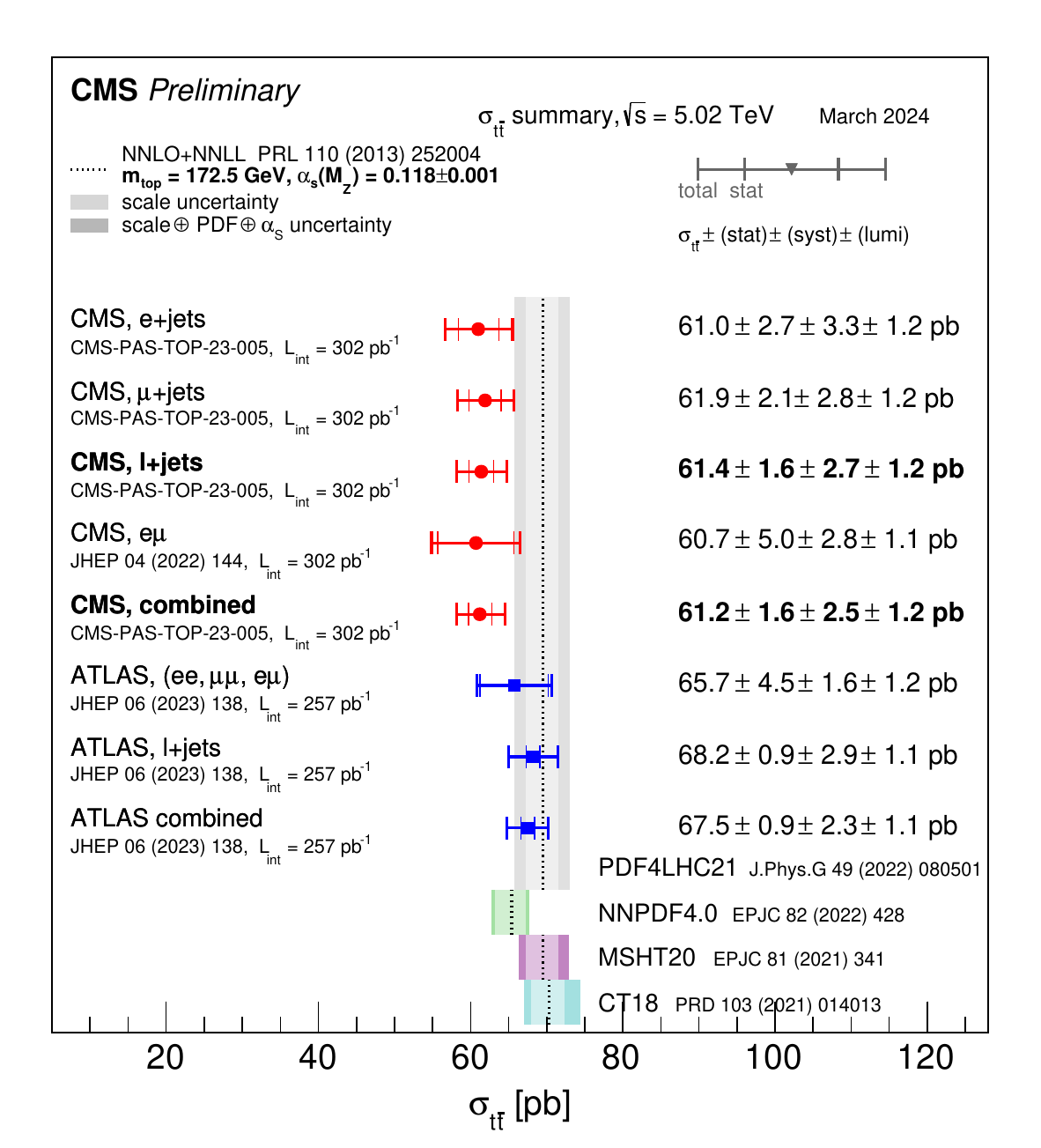}}
\end{minipage}
\hfill
\begin{minipage}{0.6\linewidth}
\centerline{\includegraphics[width=1\linewidth]{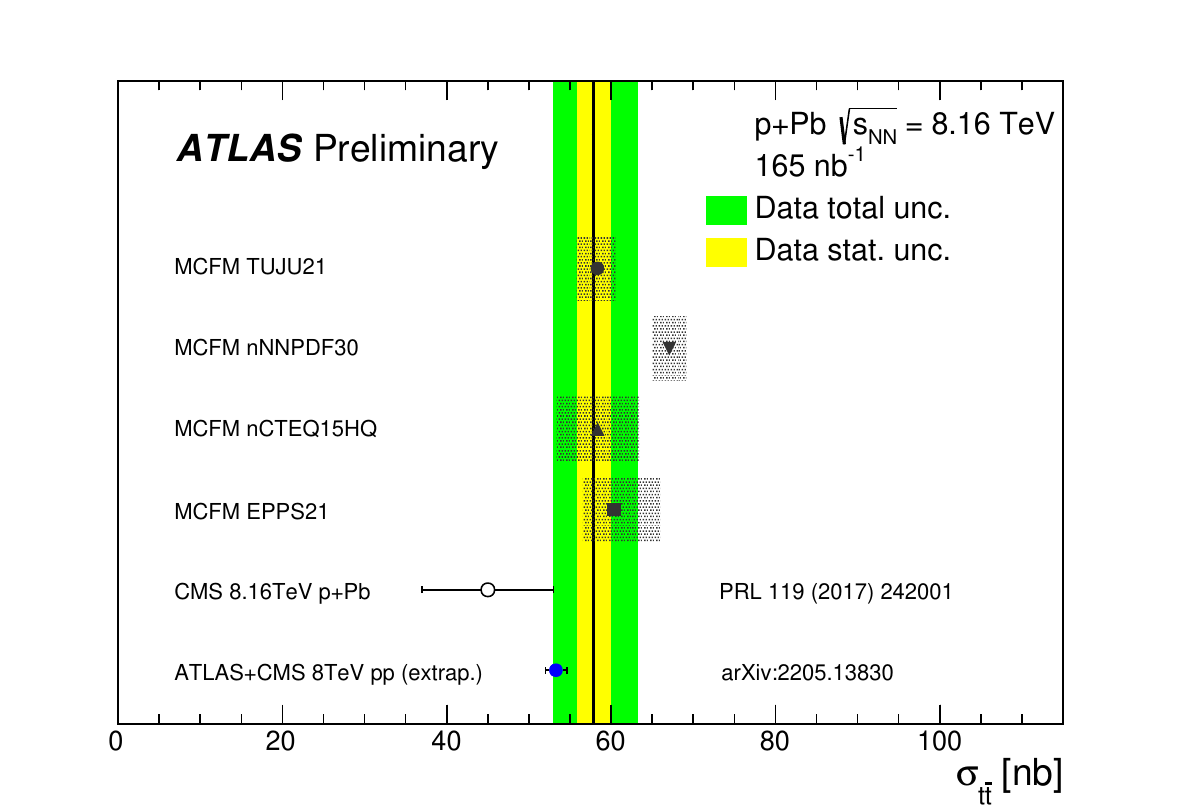}}
\end{minipage}
\caption[]{Summary of inclusive $\mathrm{t\overline{t}}$ cross section measurements at $\sqrt{s} = 5.02 \, \mathrm{TeV}$ at the LHC and their comparison to the prediction (left)~\cite{cmstt5}, and comparison of the measured $\mathrm{t\overline{t}}$ cross section in proton-lead collisions by the ATLAS Collaboration with previous measurements and predictions using different nuclear parton density functions (right)~\cite{atlasttlead}.}
\label{fig:topincl}
\end{figure}

With around $140\,\mathrm{fb}^{-1}$ of proton-proton collision data collected at a center-of-mass energy of $\sqrt{s} = 13 \, \mathrm{TeV}$ by the ATLAS~\cite{atlas,atlasrun3} and CMS~\cite{cms,cmsrun3} experiments during Run 2 of the LHC the production cross section of $\mathrm{t\overline{t}}$ can be precisely measured in all different final states. A measurement by the ATLAS Collaboration in the dilepton channel achieves the highest precision of around 1.8\%~\cite{atlastt13}. The $\mathrm{t\overline{t}}$ production cross section has also been measured during a low-intensity run of the LHC during Run 2 at $\sqrt{s} = 5.02 \, \mathrm{TeV}$, for which the CMS Collaboration has released a new measurement in the lepton+jets final state~\cite{cmstt5}. The measured cross section of $\sigma_{\mathrm{t\overline{t}}} = 61.4 \pm 1.6 \,(\mathrm{stat}) ^{+2.7}_{-2.6} \, (\mathrm{syst}) \pm 1.2 \,(\mathrm{lumi}) \,\mathrm{pb}$ is below the prediction, but still compatible with it within two standard deviations. In addition the result is combined with an already existing measurement in the dilepton final state. The results are summarized in Fig.~\ref{fig:topincl} (left).
\\
Both the ATLAS and CMS Collaborations have measured the $\mathrm{t\overline{t}}$ cross section at the currently still ongoing Run 3 of the LHC at $\sqrt{s} = 13.6 \, \mathrm{TeV}$. While the ATLAS Collaboration measures the cross section together with the cross section for Z boson production in the entire 2022 data set~\cite{atlastt136}, the CMS Collaboration uses only a fraction of the 2022 data set~\cite{cmstt136}. Both experiments already achieve a precision below 4\% and measure a lower cross section compared to the prediction, but are both compatible with the prediction within the uncertainties.\\
In addition to proton-proton collisions, the $\mathrm{t\overline{t}}$ production cross section can also be measured in proton-lead collision, as carried out by the ATLAS Collaboration recently~\cite{atlasttlead}. By combining multiple different final states the production of $\mathrm{t\overline{t}}$ was observed with a high significance and the sensitivity of the measured value allows to distinguish between different predictions of nuclear parton density functions, as indicated on Fig.~\ref{fig:topincl} (right).

\section{Differential top quark cross sections}
\begin{figure}
\begin{minipage}{0.5\linewidth}
\centerline{\includegraphics[width=1\linewidth]{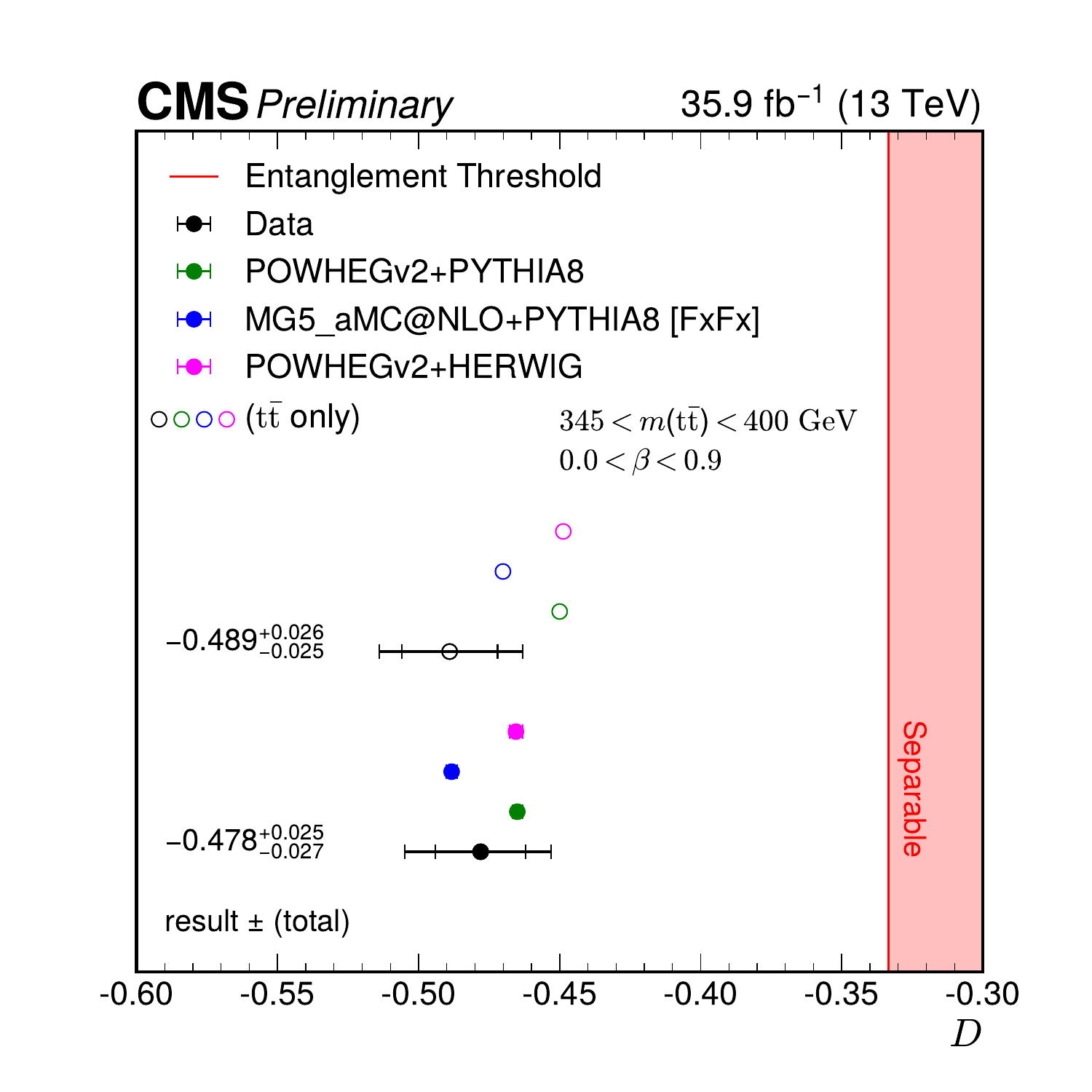}}
\end{minipage}
\hfill
\begin{minipage}{0.5\linewidth}
\centerline{\includegraphics[width=1\linewidth]{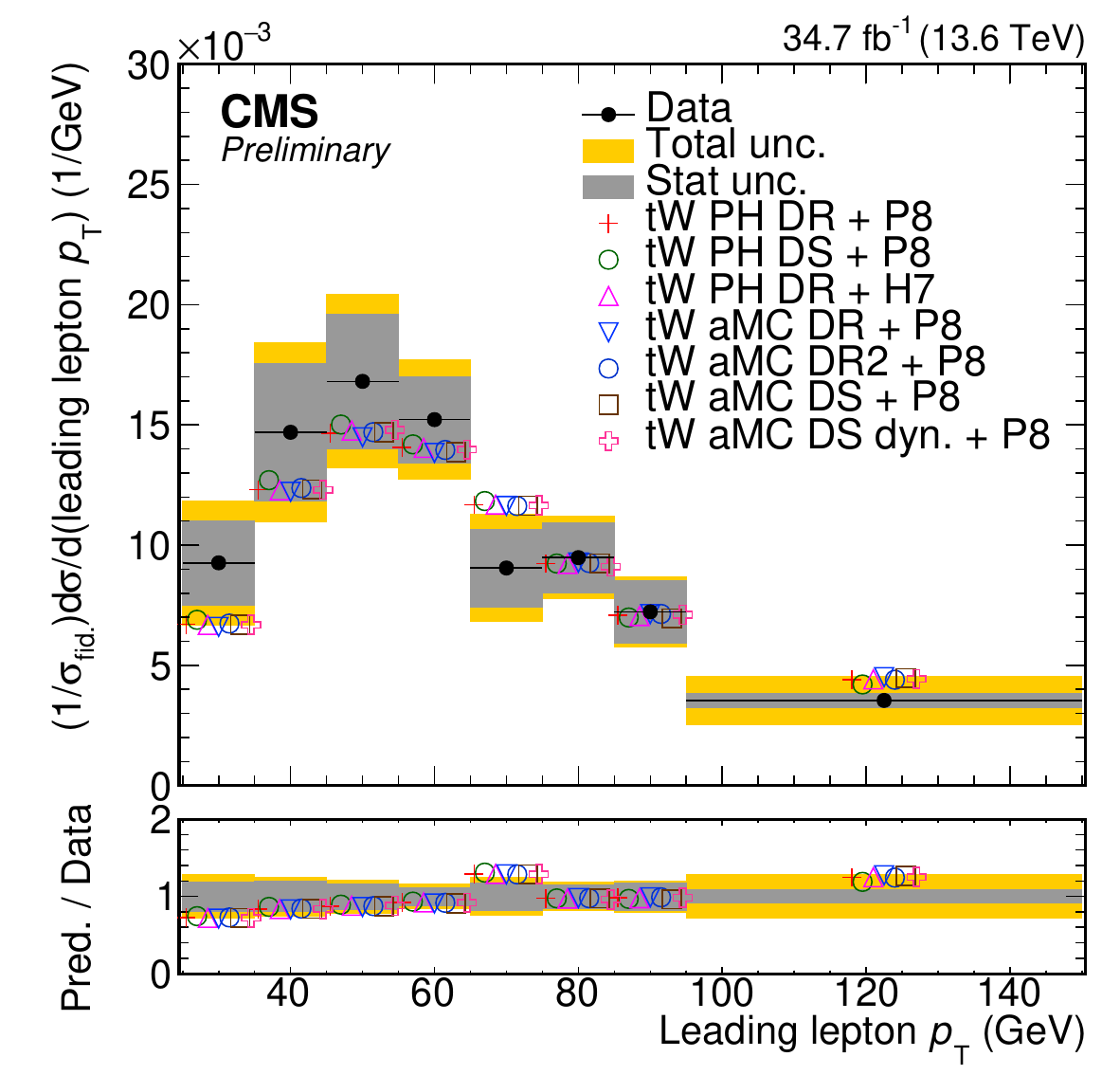}}
\end{minipage}
\caption[]{Observed and predicted values for the entanglement proxy $D$ (left)~\cite{cmsent}, which results in the observation of top quark entanglement in $\mathrm{t\overline{t}}$ events, and the unfolded distribution of the lepton with the highest transverse momentum in the differential tW measurement at Run 3 (right)~\cite{cmstw13}.}
\label{fig:topdiff}
\end{figure}

The vast amount of $\mathrm{t\overline{t}}$ produced at the LHC, especially at $\sqrt{s} = 13 \, \mathrm{TeV}$, allows not only to measure the cross section inclusively, but also differentially, i.e., with respect to one or more kinematic variables or properties of the collision event. These measurements have been performed by the ATLAS and CMS Collaborations in the lepton+jets~\cite{atlastt13dif} and dilepton~\cite{cmstt13dif} decay channels, respectively. The measured distributions are unfolded to particle or parton level and then compared to various different theoretical predictions. It is found that, e.g., the description of the top quark transverse momentum at higher values is significantly improved at next-to-next-to-leading order in quantum chromodynamics in comparison to next-to-leading order. On the other hand, most predictions fail to properly describe the invariant mass distribution of the $\mathrm{t\overline{t}}$ system for masses at the threshold region.\\
This effect has also been studied in a new differential measurement by the CMS Collaboration, where the spin correlation of the charged leptons from $\mathrm{t\overline{t}}$ decays was analyzed~\cite{cmsent}. By measuring the helicity angle of the two leptons it was found that an improved description of the data can be achieved by including a potential bound $\mathrm{t\overline{t}}$ state, called toponium ($\eta_{\mathrm{t}}$), to the signal definition. In addition the quantum entanglement of top quarks in $\mathrm{t\overline{t}}$ production was observed with a significance above $5\sigma$, as indicated in Fig.~\ref{fig:topdiff} (left).
\\
The first differential top quark measurement at Run 3 of the LHC has been performed by the CMS Collaboration by measuring tW-associated single top quark production with the full 2022 data set~\cite{cmstw13}. Events are required to contain an electron and muon of opposite charge and exactly one jet, which is identified as originating from a bottom quark. The resulting distributions are unfolded to particle level and compared to various predictions, exemplary shown in Fig.~\ref{fig:topdiff} (right) for the lepton in the event with the highest transverse momentum. The cross section has also been measured inclusively and the obtained value is well in agreement with the prediction.

\section{Top quark mass}
\begin{figure}
\begin{minipage}{0.55\linewidth}
\centerline{\includegraphics[width=1\linewidth]{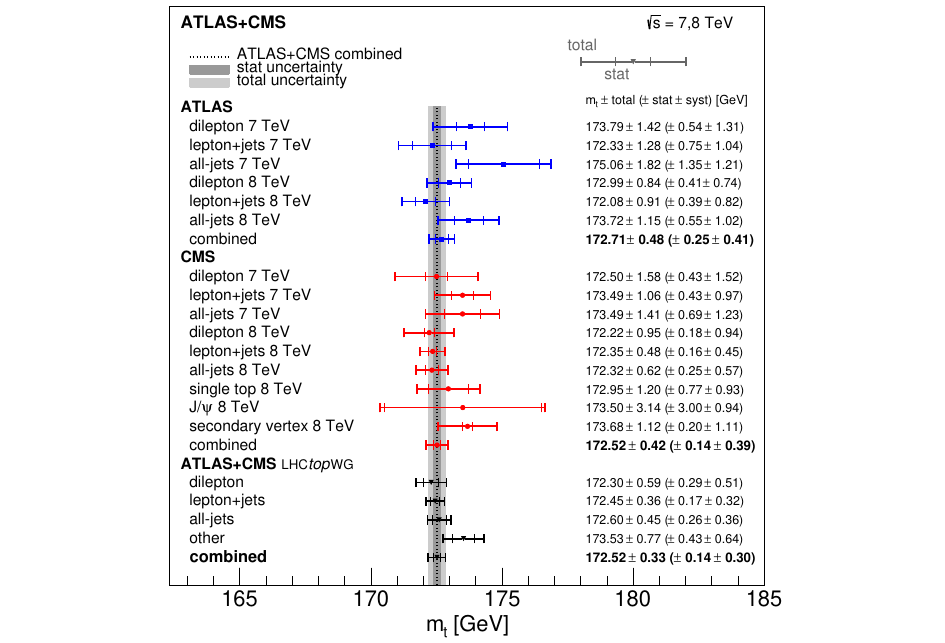}}
\end{minipage}
\hfill
\begin{minipage}{0.45\linewidth}
\centerline{\includegraphics[width=1\linewidth]{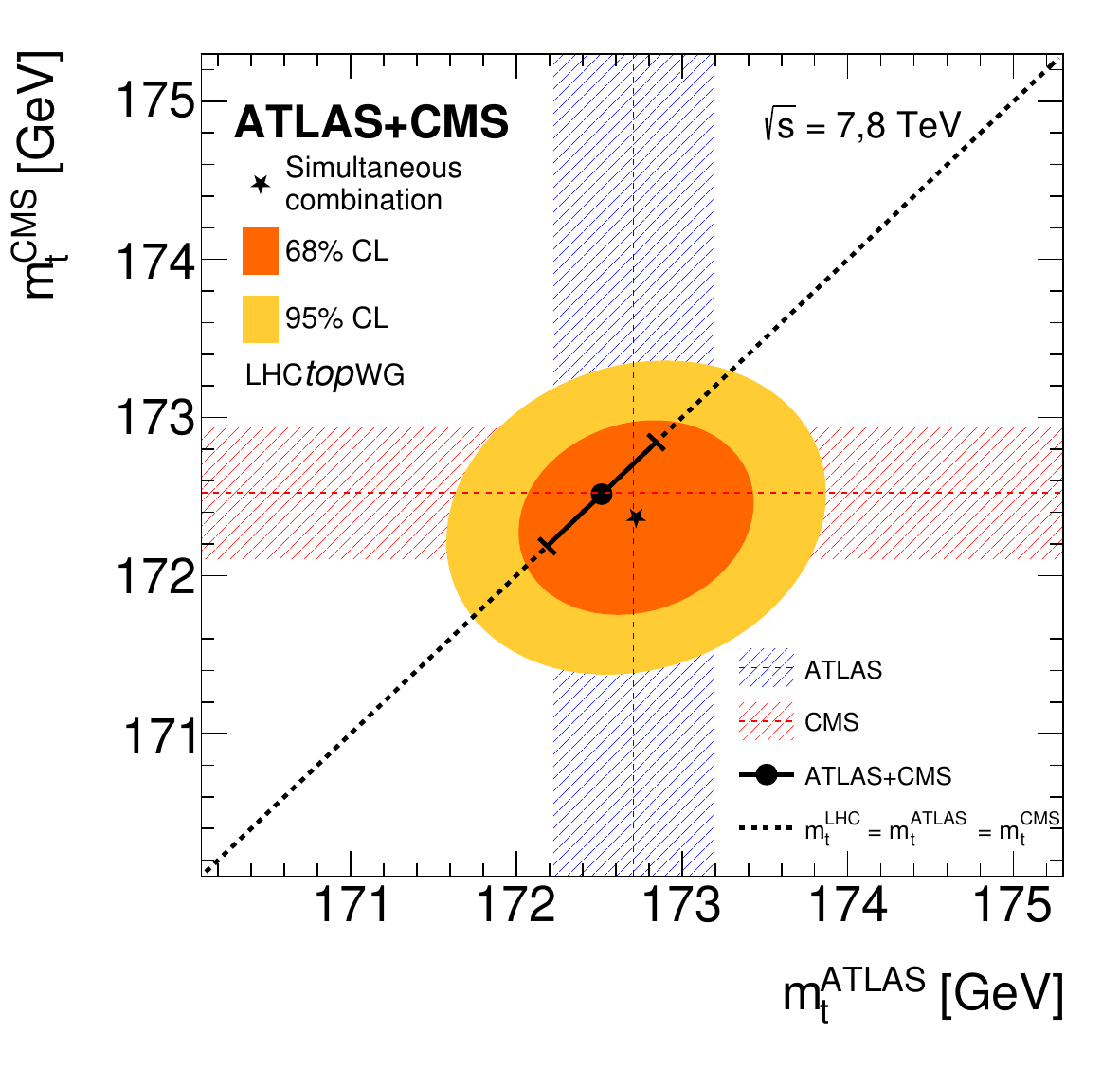}}
\end{minipage}
\caption[]{The input measurements and all different combined values of the $m_{\mathrm{t}}$ combination (left) and the relation between the combined values from the individual combinations of the measurements from the ATLAS and CMS Collaborations, as well as the total combined value at the LHC (right)~\cite{masscomb}.}
\label{fig:topmass}
\end{figure}

The top quark mass ($m_{\mathrm{t}}$) is in an important free parameter in the SM. With its value close to the electroweak scale and entering loop contributions of other processes it is of great importance to measure $m_{\mathrm{t}}$ as precisely as possible. The top quark mass can be measured indirectly through top quark cross sections, where the results depend on the value of $m_{\mathrm{t}}$ realized in nature, or directly by measuring the reconstructed $m_{\mathrm{t}}$ or equivalent quantities.\\
The CMS Collaboration has measured a top quark mass of $m_{\mathrm{t}} = 173.06 \pm 0.84\,\mathrm{GeV}$ in $\mathrm{t\overline{t}}$ events with all-jet final states at $\sqrt{s} = 13 \, \mathrm{TeV}$, where the decay products of each top quark are boosted in a way that they can be clustered all together in a single, large-radius jet~\cite{cmsmassqq}. A measurement by the ATLAS Collaboration in the dilepton decay channels obtains a value of $m_{\mathrm{t}} = 172.21 \pm 0.20 \,(\mathrm{stat})\pm 0.67 \, (\mathrm{syst}) \pm 0.39 \,(\mathrm{recoil}) \,\mathrm{GeV}$~\cite{atlasmassll}. In this analysis the mass is extracted from a template fit to the $m_{\ell\mathrm{b}}$ distribution without relying on the difficult full reconstruction of the $\mathrm{t\overline{t}}$ system in this final state. The single most precise determination of $m_{\mathrm{t}}$ is performed by the CMS Collaboration in an analysis where the full $\mathrm{t\overline{t}}$ event kinematic in lepton+jets final states is taken into account~\cite{cmsmasslj}. A profile likelihood method is used to measure a value of $m_{\mathrm{t}} = 171.77 \pm 0.37\,\mathrm{GeV}$.\\
In an effort to obtain an even more precise determination of $m_{\mathrm{t}}$ the ATLAS and CMS Collaborations have combined their previous measurements during Run 1 of the LHC at $\sqrt{s} = 7$ and $8 \, \mathrm{TeV}$~\cite{masscomb}. In total 6 measurements from the ATLAS experiment and 9 measurements from the CMS experiments are included in this combination, which utilizes the best linear unbiased estimate (BLUE) technique~\cite{blue}. An overview of all input measurements and resulting combined values is provided in Fig.~\ref{fig:topmass} (left), while Fig~\ref{fig:topmass} (right) highlights the relation between the individual combined values from both experiments and the total combined value. When combining all 15 measurements a value of $m_{\mathrm{t}} = 172.52 \pm 0.33 \, \mathrm{GeV}$ is achieved, the most precise determination of $m_{\mathrm{t}}$ to date.

\section{Conclusion}
Almost thirty years after its first discovery at the Tevatron collider the top quark remains a subject of current high energy physics research. With progress in experimental techniques and theoretical calculations in recent years, as well as the large amount of data collected at the LHC, properties of the top quark, such as its inclusive production cross section and mass, can be measured with an accuracy of a few percent. With the help of the ongoing Run 3 of the LHC this precision can be further increased, especially with differential measurements, which will push the boundaries of our understanding of the top quark and its associated phenomena.

\end{document}